\documentclass[11pt,a4paper]{article}
\usepackage[T1]{fontenc}
\usepackage[utf8]{inputenc}
\usepackage{lmodern}
\usepackage[margin=2.5cm]{geometry}
\usepackage{amsmath}
\usepackage{graphicx}
\usepackage{booktabs}
\usepackage{array}
\usepackage[font=small,labelfont=bf]{caption}
\usepackage{microtype}
\usepackage[section]{placeins}
\usepackage[hidelinks]{hyperref}
\hypersetup{pdftitle={Blind tests of a simple-body model of submarine target strength on open benchmarks},
            pdfauthor={Sunil Tyagi}}

\newcommand{\dB}{\,dB}
\newcommand{\Req}{R_{\mathrm e}}
\newcommand{\rhoF}{\rho_{\mathrm F}}
\title{Blind tests of a simple-body model of submarine target strength on open benchmarks}
\author{Sunil Tyagi\\[2pt]
{\normalsize Oravont Systems LLP, Noida, India}\\
{\normalsize\texttt{suniltyagi@tyagination.com}}}
\date{19 November 2026}

\begin{document}
\maketitle

\begin{abstract}
\noindent Finite-element and boundary-element models of a submarine's sonar echo need a mesh and a solver, and their cost grows quickly with frequency.
SSR-Sim takes an analytical route. The hull is cut into a few simple bodies, the echo of each is a closed-form result of geometrical acoustics
or of the Kirchhoff approximation, the surface enters through a plane-wave reflection coefficient, and the echoes add in energy.
This paper describes the method, including the reduction of a hull and its sail to simple bodies, and tests it on three open cases.
(1)~The reflection coefficient of a 50\,mm rubber tile on steel plating, computed with finite elements by Yang et~al.,
is reproduced to 0.02 rms or better at incidence angles from $0^\circ$ to $75^\circ$ once the layer is allowed to refract.
(2)~For the rigid full-scale BeTSSi submarine at 1\,kHz, a prediction fixed before the reference was opened agrees with the
boundary-element results of Ven{\aa}s and Kvamsdal to within 0.7\dB{} in every aspect sector from the bow to $150^\circ$;
in the stern sector it is 5\dB{} high.
(3)~For a 1:15 BeTSSi model with 10\,mm tiles measured in a lake, a second blind prediction is within $-3.1$, $+1.4$ and $-0.5$\dB{}
of the measured level over aspects $0$--$120^\circ$ at 5, 15 and 20\,kHz, against $-3.1$, $+3.1$ and $+3.5$\dB{} for the
Kirchhoff--finite-element computation published with the measurement.
At 10\,kHz, on a resonance of the tile, the prediction is 6.9\dB{} low, and in the bow sector it is 7\dB{} or more low at every frequency.
The tests cover the specular component and the coating model only.
The failures show what such a model lacks: edge diffraction, and the returns that set the off-beam floor of a real model.
\end{abstract}


\section{Introduction}

The target strength (TS) of a submarine decides how far away an active sonar can detect it, and it is one of the few
signature quantities a designer can change through the hull form and its coatings.
Three families of method predict it.
Full-wave methods, boundary and finite elements, solve the scattering problem and are the reference where they can be afforded~\cite{venas2020};
their cost grows quickly with frequency.
Kirchhoff, or physical-optics, methods integrate over a facet mesh of the lit surface~\cite{abawi2016,fawcett2001,yang2024}.
Component models are older and simpler than both: the hull is a short list of spheres, cylinders, cones and plates,
each with a closed-form echo, and the echoes are added~\cite{urick1983}.
A component model needs no mesh and no solver.
It gives a level in a fraction of a millisecond, and every decibel of its output can be traced to one body.
That suits early design, coating trade studies and any use that needs thousands of aspects and frequencies.

The weakness of component models is trust. Their forms are high-frequency approximations, the cutting of a hull into bodies is a
judgement, and comparisons with open data are seldom published.
The BeTSSi benchmark submarine~\cite{nell2003,ostberg2016} now offers such data: a public geometry, full-wave results for the rigid hull~\cite{venas2020,nird2019},
and a lake measurement of a tile-coated scale model~\cite{yang2024}.

This paper describes one component model, SSR-Sim~\cite{tyagi2026}, and tests it on three open cases.
SSR-Sim is analytical throughout: no finite-element or boundary-element model of the hull is made at any stage.
The cases are the reflection of a rubber tile, the rigid full-scale BeTSSi
submarine, and the coated 1:15 scale model. The two hull cases are blind: the prediction was fixed and recorded before the reference curve was opened.
Section~\ref{sec:model} gives the method: the theory behind each term, the way a hull is cut into simple bodies, and the treatment of the sail,
in enough detail for the predictions to be recomputed. Sections~\ref{sec:protocol} to~\ref{sec:scale} give the tests.

\section{The method}\label{sec:model}

\subsection{An analytical model in place of a numerical one}

A full-wave model solves the wave equation on a mesh of the wetted surface. A facet Kirchhoff model integrates over such a mesh.
SSR-Sim uses neither a mesh nor a solver. It rests on four approximations, all of them classical.
\begin{enumerate}\setlength{\itemsep}{0pt}
\item \emph{High frequency.} Every reflecting surface is large against the wavelength.
Its echo then comes from the neighbourhood of the points where its normal points at the sonar, the specular points.
\item \emph{Locally plane reflection.} At a specular point the surface reflects as an infinite plane of the same construction would at the same angle of incidence.
The construction (plating, coating, what lies behind) enters only through a plane-wave reflection coefficient $\beta$.
\item \emph{Simple bodies.} The hull is replaced by a short list of bodies: sphere, spheroid, doubly curved patch, cylinder, cone, plate, disc.
For each of them approximations 1 and 2 give the echo in closed form.
\item \emph{Energy summation.} The echoes of the bodies are added in energy, without phase.
\end{enumerate}
The water is unbounded, source and receiver are at the same point, and the pulse is long enough for the echoes of all the bodies to overlap.
The tests in this paper are in the horizontal plane.

The price of these approximations is known in advance. The specular model has no diffraction by edges, tips and junctions, no shadowing of one body by another,
no multiple reflection between bodies, and no response of the structure other than through $\beta$.
Sections~\ref{sec:rigid} and~\ref{sec:scale} measure what that costs.

\subsection{Equivalent radius and the energy sum}

SSR-Sim expresses every echo as an equivalent radius $\Req$, the radius of the large rigid sphere that would return the same echo.
Target strength follows as
\begin{equation}
  \mathrm{TS} = 20\lg\frac{\Req}{2}\quad\text{(dB re 1 m)}.
\end{equation}
The equivalent radii of the bodies add in energy,
\begin{equation}
  \Req^2=\sum_i R_i^2 .
\end{equation}
The sum without phase is deliberate. Bodies tens of wavelengths apart change their relative phase through many cycles within a degree of aspect,
and on a real boat those phases depend on details that no drawing fixes.
The result is therefore a phase-averaged level: it reproduces the mean of an interference pattern, not its lobes.

\subsection{Specular reflection}\label{sec:specular}

The specular term of body $i$ at aspect $\alpha$ is
\begin{equation}\label{eq:term}
  R_i(\alpha) = R_{0,i}\,\varphi_i(\theta)\,\beta_i(\theta),
\end{equation}
where $\theta$ is the angle between the line of sight and the body's normal (zero for a curved body, whose specular point faces the sonar), $R_{0,i}$ is the equivalent radius of the rigid body at normal incidence,
$\varphi_i$ is its directivity and $\beta_i$ its reflection coefficient. Table~\ref{tab:forms} lists the forms of $R_0$.
They come from two standard evaluations of the Kirchhoff integral, in which the pressure on the lit surface is taken as that on its tangent plane~\cite{bowman1969,crispin1965}.
Most of them are tabulated for sonar targets by Urick~\cite{urick1983}.

\paragraph{Curved bodies: geometrical acoustics.} For a smooth convex surface the integral is evaluated by stationary phase.
Only the neighbourhood of the specular point contributes, and the surface enters through its two principal radii of curvature there:
\begin{equation}\label{eq:curved}
  R_0=\sqrt{\rho_1\rho_2}.
\end{equation}
A sphere of radius $a$ gives $R_0=a$, which is the definition of the equivalent radius.
A spheroid with semi-axis $A$ along its axis and $B$ across it, seen at angle $\alpha$ from the axis, gives
\begin{equation}\label{eq:spheroid}
  R_0=\frac{AB^2}{A^2\cos^2\alpha+B^2\sin^2\alpha},
\end{equation}
which is~\eqref{eq:curved} at its specular point. A curved body has no directivity function.
It returns~\eqref{eq:curved} at every aspect within the range of its surface normals and nothing outside that range.
For a patch or the front of a foil the term is halved at the end of that range, where the specular point is on the edge of the body and only half of the first Fresnel zone lies on it.

\paragraph{Ruled bodies: physical optics.} For a body with straight generators (a plate, a cylinder, a cone) the integral along the generator is done exactly,
and the integral around a cylinder or cone by stationary phase. At normal incidence to the generator this gives the forms of Table~\ref{tab:forms}:
$R_0=L\sqrt{2a/\lambda}$ for a cylinder of length $L$ and radius $a$, and $R_0=2LH/\lambda$ for a plate of sides $L$ and $H$.
Away from the normal the integral along the generator gives the lobe function
\begin{equation}\label{eq:lobe}
  \varphi=\left|\frac{\sin x}{x}\right|,\qquad x=kL\sin\theta,
\end{equation}
where $k=2\pi/\lambda$ and $L$ is the extent of the body in the plane of the sweep: the generator of a cylinder or cone, the side of a plate, the diameter of a disc.
The slowly varying obliquity factors of the Kirchhoff integral ($\cos\theta$ for a plate, $\sqrt{\cos\theta}$ for a cylinder) are not carried.
They matter only far from the normal, where the approximation itself is poor.
A ruled body returns nothing beyond $90^\circ$ from its normal, and a cone or frustum of half-angle $\psi$ is counted only at aspects more than $\psi$ from its axis.

\paragraph{A cylinder of any section.} The stationary-phase step does not need a circular section.
A cylinder of any smooth convex section returns as a circular cylinder whose radius is the section's radius of curvature $\rho$ at the specular point:
\begin{equation}\label{eq:anycyl}
  R_0=L\sqrt{2\rho/\lambda}.
\end{equation}
This is the form used for a sail and for other foils (Section~\ref{sec:sail}).

\begin{table}[!htb]
\centering\small
\caption{Specular forms: the equivalent radius $R_0$ of a rigid body at normal incidence. GA is geometrical acoustics (stationary phase about the specular point);
PO is physical optics (the Kirchhoff integral along the generator). A complete cone is a frustum with $a_2=0$.}\label{tab:forms}
\renewcommand{\arraystretch}{1.35}
\begin{tabular}{@{}p{5.6cm}p{4.2cm}p{4.0cm}l@{}}
\toprule
Body & $R_0$ & Directivity & Basis\\
\midrule
Sphere, radius $a$ & $a$ & none, within its normals & GA\\
Doubly curved patch, principal radii $\rho_1$, $\rho_2$ & $\sqrt{\rho_1\rho_2}$ & none, within its normals & GA\\
Spheroid, semi-axes $A$ (axis) and $B$, at angle $\alpha$ from its axis & $AB^2/(A^2\cos^2\alpha+B^2\sin^2\alpha)$ & none, within its sector & GA\\
Cylinder, length $L$, radius $a$ & $L\sqrt{2a/\lambda}$ & lobe~\eqref{eq:lobe}, its length & PO\\
Cylinder of any convex section, local radius $\rho$ & $L\sqrt{2\rho/\lambda}$ & as the cylinder & PO\\
Frustum, end radii $a_1>a_2$, half-angle $\psi$ & $\dfrac{2\sqrt2}{3}\,\dfrac{a_1^{3/2}-a_2^{3/2}}{\sqrt{\lambda}\,\sin\psi\,\sqrt{\cos\psi}}$ & lobe~\eqref{eq:lobe}, its generator & PO\\
Plate, sides $L$ and $H$ & $2LH/\lambda$ & lobe~\eqref{eq:lobe}, its side in the sweep & PO\\
Disc, radius $a$ & $2\pi a^2/\lambda$ & lobe~\eqref{eq:lobe}, its diameter & PO\\
\bottomrule
\end{tabular}
\end{table}

\subsection{Range}\label{sec:range}

The ruled forms of Table~\ref{tab:forms} assume that the whole body reflects in phase.
At range $r$ that holds while the body is shorter than the first Fresnel zone, whose diameter is
\begin{equation}\label{eq:fresnel}
  2\rhoF=\sqrt{2r\lambda}.
\end{equation}
A longer body returns from that zone only, and its echo stops growing with its length:
a long cylinder gives $R_0=\sqrt{ar}$, a long strip of height $H$ gives $R_0=H\sqrt{2r/\lambda}$, and a plate larger than the zone in both directions gives $R_0=r$, the image of the source.
SSR-Sim tests every body against $2\rhoF$. For a long body it uses the long form inside the main lobe, $\theta\le\arcsin(\lambda/2L)$, and the short form with its lobe function outside it, where the returns come from the ends of the body.
A curved patch enters~\eqref{eq:curved} with each radius reduced to $\rho/(1+\rho/r)$.
The long forms matter for a long hull at short range or high frequency, and in a test tank (Section~\ref{sec:scale}).

\subsection{The reflection coefficient}\label{sec:beta}

$\beta$ is the modulus of the plane-wave pressure reflection coefficient of the surface at the local angle of incidence. On a coated surface that angle is limited to $85^\circ$.

\paragraph{Rigid surface.} $\beta=1$.

\paragraph{Coated plating.} A coating enters through the input impedance $Z_{\mathrm{in}}$ of the coated surface, which gives
\begin{equation}
  \beta = \left|\frac{Z_{\mathrm{in}}\cos\theta-\rho_0c_0}{Z_{\mathrm{in}}\cos\theta+\rho_0c_0}\right| .
\end{equation}
For a homogeneous viscoelastic layer of thickness $h$, density $\rho$ and complex plane-wave modulus $M(1-\mathrm i\eta)$ the layer is treated as a fluid layer~\cite{brekhovskikh1980}.
With the time dependence $\mathrm e^{-\mathrm i\omega t}$,
\begin{equation}\label{eq:layer}
  Z_{\mathrm{in}} = Z\,\frac{Z_{\mathrm b}-\mathrm i Z\tan k_zh}{Z-\mathrm i Z_{\mathrm b}\tan k_zh},\qquad
  k_z=k_{\mathrm p}\cos\theta_{\mathrm p},\quad Z=\frac{\rho c_{\mathrm p}}{\cos\theta_{\mathrm p}},\quad \sin\theta_{\mathrm p}=\frac{c_{\mathrm p}}{c_0}\sin\theta ,
\end{equation}
where $c_{\mathrm p}=\sqrt{M(1-\mathrm i\eta)/\rho}$ and $k_{\mathrm p}=\omega/c_{\mathrm p}$.
The plate under the layer acts as a mass $m$ per unit area, so the backing impedance is $Z_{\mathrm b}=-\mathrm i\omega m$ with air behind the plate
and $Z_{\mathrm b}=-\mathrm i\omega m+\rho_0c_0/\cos\theta$ with water behind it.
Setting $\theta_{\mathrm p}=0$ gives a locally reacting layer. Shear waves in the layer are neglected.

\subsection{What is not tested here}\label{sec:other}

SSR-Sim also has terms for two further returns of a double-hull boat: diffuse scattering from the structure between the hulls,
and non-specular re-radiation from stiffened plating~\cite{tyagi2026}.
Neither is exercised below: a rigid body and a single frameless shell have no such structure, and no open benchmark has it.
They are outside this paper, and so is the reflection of bare plating, which no case here uses.

\subsection{Cutting a hull into bodies}\label{sec:cut}

\begin{figure}[!htb]
\centering
\includegraphics[width=\textwidth]{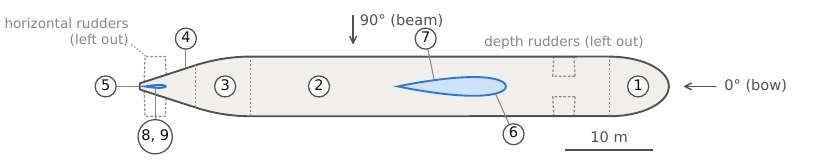}
\caption{The BeTSSi hull in plan at the level of its axis, cut into the bodies of Table~\ref{tab:bodies}: 1~half spheroid, 2~cylinder, 3~doubly curved patch, 4~frustum, 5~disc,
6 and~7 the front and sides of the sail, 8 and~9 the front and sides of the vertical rudders. The sail and the vertical rudders are shown by their sections.}\label{fig:bodies}
\end{figure}

The hull is cut where its form changes: where a spheroid meets a cylinder, a cylinder a curved run, a curved run a cone.
Table~\ref{tab:parts} gives the usual assignments, and Fig.~\ref{fig:bodies} shows them on the BeTSSi hull. Three points need care.
\begin{itemize}\setlength{\itemsep}{0pt}
\item Only a surface with a normal in the plane of the sweep returns a specular echo. In the horizontal plane a flat deck, the top of a sail and the horizontal planes have none, and they are left out.
\item A curved run between a cylinder and a cone, as at the stern, is a doubly curved patch. On a surface of revolution the principal radii at a point are the radius of the meridian and $r/\cos\psi$,
where $r$ is the distance of the point from the axis and $\psi$ the angle between its normal and the plane normal to the axis.
The patch returns at the aspects between the normal of the cylinder and that of the cone.
\item Each body is entered once and returns at every aspect its form allows. The model does not hide one body behind another.
\end{itemize}

\begin{table}[!htb]
\centering\small
\caption{Approximation of the parts of a submarine by simple bodies, for a sweep in the horizontal plane. $\psi$ is the half-angle of the stern cone.}\label{tab:parts}
\begin{tabular}{@{}p{4.3cm}p{5.2cm}p{5.5cm}@{}}
\toprule
Part & Simple body & Where it returns\\
\midrule
Bow & half spheroid, or a spherical cap & from ahead to the beam, without lobes\\
Parallel body & circular cylinder & a flash on the beam, side lobes elsewhere\\
Curved run to the stern & doubly curved patch & between the beam and $90^\circ+\psi$\\
Stern cone & cone or frustum & a flash at $90^\circ+\psi$, side lobes elsewhere\\
Blunt end & disc & about $180^\circ$\\
Sail, vertical rudders: front & vertical cylinder of the section's local radius, eq.~\eqref{eq:anycyl} & from ahead to the beam, without lobes\\
Sail, vertical rudders: sides & flat plates & a flash just aft of the beam, side lobes elsewhere\\
Deck, top of the sail, horizontal planes & none & no specular echo in this plane\\
\bottomrule
\end{tabular}
\end{table}

\subsection{The sail}\label{sec:sail}

\begin{figure}[!htb]
\centering
\includegraphics[width=\textwidth]{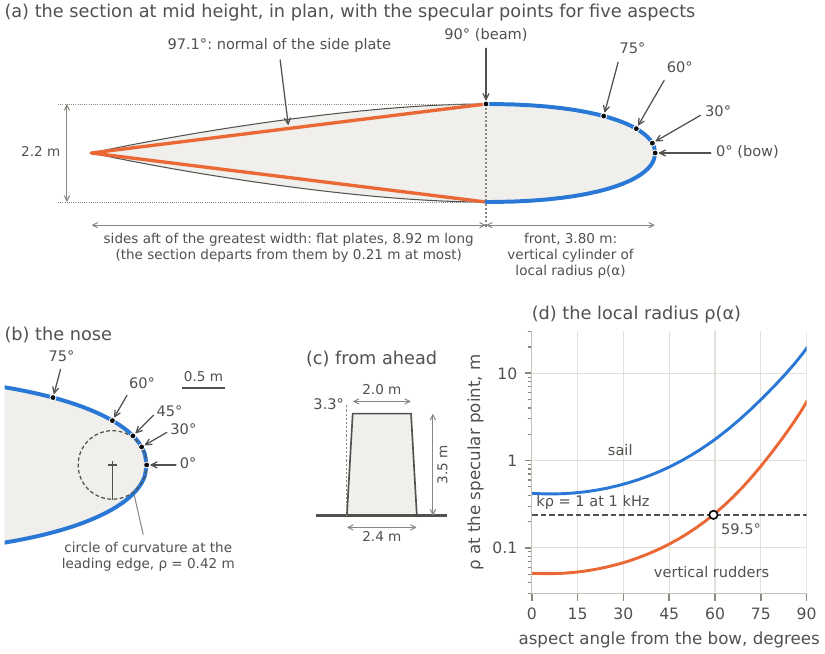}
\caption{The model of the BeTSSi sail. (a)~The section at mid height. The front (blue) is a vertical cylinder whose radius is the section's local radius of curvature;
each side (orange) is a flat plate. (b)~The nose, with the circle of curvature at the leading edge. (c)~The sail from ahead: its walls lean inwards.
(d)~The radius of curvature at the specular point against aspect, for the sail and for the vertical rudders; a curved front is counted where $k\rho\ge1$.}\label{fig:sail}
\end{figure}

A sail is neither a cylinder nor a plate. Its section in plan is a foil: sharply curved at the leading edge, nearly flat along the sides, and ending in an edge.
It is modelled in two parts, divided at the greatest width of the section (Fig.~\ref{fig:sail}). The BeTSSi sail serves as the example;
its sections are NACA foils~\cite{venas2020}, and the section at mid height, of chord 12.65\,m and width 2.2\,m, is used for the whole height $H=3.5$\,m.

\paragraph{Front.} From the leading edge to the greatest width the section turns through $90^\circ$ on each side.
For every aspect from $0^\circ$ to $90^\circ$ there is therefore one point on the near side whose normal points at the sonar.
By~\eqref{eq:anycyl} the front returns as a vertical cylinder of height $H$ whose radius is the radius of curvature of the section at that point:
\begin{equation}\label{eq:front}
  R_{\mathrm{front}}(\alpha)=H\sqrt{\frac{2\rho(\alpha)}{\lambda}}\;\varphi_{\mathrm v}\,\beta,\qquad
  \rho=\frac{(1+y'^2)^{3/2}}{|y''|}\ \ \text{where}\ \ y'=\cot\alpha .
\end{equation}
Here $y(u)$ is the half-width of the section at distance $u$ from the leading edge, and $\varphi_{\mathrm v}$ is a factor for the lean of the walls, given below.
There is no lobe function in azimuth, because the generator is vertical and the sweep is horizontal.
For a NACA section of chord $l$ and thickness ratio $t$ the half-width is $y=5tl\,(a_0\sqrt{\xi}+a_1\xi+a_2\xi^2+a_3\xi^3+a_4\xi^4)$ with $\xi=u/l$,
and the radius at the leading edge is $\rho(0)=12.5\,a_0^2t^2l$.
For the BeTSSi sail $\rho$ is 0.42\,m at the leading edge, 0.54\,m at $30^\circ$, 1.75\,m at $60^\circ$ and 19.3\,m abeam (Fig.~\ref{fig:sail}d).
The specular point stays within 0.42\,m of the leading edge at every aspect up to $60^\circ$:
seen from within $60^\circ$ of the bow, the echo of a sail 12.65\,m long comes from its first half metre.

\paragraph{Validity of the front.} Equation~\eqref{eq:anycyl} needs $k\rho\gg1$.
For a rigid circular cylinder the exact series solution~\cite{bowman1969} stays within $+0.9$ and $-2.5$\dB{} of the form for $ka\ge0.6$,
and falls away below that: 4\dB{} under it at $ka=0.4$ and 13\dB{} under it at $ka=0.2$.
A curved front is therefore counted only where $k\rho\ge1$.
At 1\,kHz the BeTSSi sail passes at every aspect ($k\rho\ge1.7$). The vertical rudders, whose leading-edge radius is 0.05\,m, pass only from $59.5^\circ$ (Fig.~\ref{fig:sail}d).
Even where the rule is met, the form is at its limit on the nose: the first Fresnel zone there has a half-width $\sqrt{\lambda\rho/2}=0.56$\,m,
as large as the region over which $\rho$ changes. The rigid benchmark is the test of it.

\paragraph{Sides.} Aft of the greatest width the section departs from the straight line that joins that point to the trailing edge by 0.21\,m at most, a seventh of a wavelength at 1\,kHz.
Each side is taken as a flat plate on that line, $L\times H=8.92\times3.5$\,m, with its normal at $\alpha_{\mathrm n}=97.1^\circ$, which is $90^\circ$ plus the taper of the section:
\begin{equation}\label{eq:side}
  R_{\mathrm{side}}(\alpha)=\frac{2LH}{\lambda}\left|\frac{\sin x}{x}\right|\varphi_{\mathrm v}\,\beta,\qquad x=kL\sin|\alpha-\alpha_{\mathrm n}| .
\end{equation}
The plate gives a flash just aft of the beam and side lobes elsewhere.
The side lobes are the returns of the two ends of the plate. One end is the junction with the front, which is not an edge at all, and the other is the trailing edge.
This is the weakest part of the model, and Section~\ref{sec:rigid} shows it.

\paragraph{Lean.} The walls of the BeTSSi sail lean inwards: the sail is 2.4\,m wide at the deck and 2.0\,m at the top, so the front leans by $\delta=3.3^\circ$ and the sides by about half that on average.
In the horizontal plane a leaning wall is seen off its normal in elevation. The lobe function in the vertical plane,
\begin{equation}\label{eq:lean}
  \varphi_{\mathrm v}=\left|\frac{\sin x_{\mathrm v}}{x_{\mathrm v}}\right|,\qquad x_{\mathrm v}=kH\sin\delta ,
\end{equation}
reduces the front by a factor 0.89 and the sides by 0.97 at 1\,kHz.

\paragraph{What is left out.} The top of the sail has no horizontal normal. The trailing edge, the tip and the junction with the deck diffract, and the model has no term for them.

\paragraph{Other foils, and a simpler variant.} Rudders and planes with a vertical span are treated as the sail is, each with its own section.
Where the section of a sail is not known, its front is taken as a half-ellipse with semi-axes $a$ along the chord and $b$ across it, cut into a few sectors of aspect.
Each sector is a vertical cylinder with the ellipse's radius of curvature at the middle of the sector, $\rho=a^2b^2/(a^2\cos^2\alpha+b^2\sin^2\alpha)^{3/2}$, halved at the two ends of the sector.
Test~3 uses this variant.

\subsection{Profiles}

SSR-Sim has two sets of options. The \emph{engineering profile} is the primary one here; it uses the lobe function and the analytic coefficients as written above.
The \emph{envelope profile} replaces each lobe function by its envelope, which is 1 inside the main lobe and falls as $1/\theta$ outside it.
It also uses rounded constants, keeps the long form of a long body at every angle, and takes a coating layer as locally reacting.
Both were computed and recorded for each blind case, with the engineering profile named as primary in advance.

\section{Protocol}\label{sec:protocol}

\paragraph{Blind predictions.} For each hull case the bodies, materials and range were fixed from the open description of the case,
the prediction was written to a file, and the file's SHA-256 was recorded before the reference curve was digitised or opened.
The comparison script checks the hash before it reads the prediction. Appendix~\ref{app:hash} lists the hashes.
The two prediction files are published~\cite{tyagi2026data}, so that anyone can check them against these hashes.
They were recorded in the author's project log and not with a public registry, so they document the order of the work without proving it.

\paragraph{What was known beforehand.} For the scale model, the source paper's text had been read for the tile data, and it states the measured ranges in words
(beam $-7$ to $-3$\dB, bow $-14$ to $-6$\dB, stern $-16$ to $-9$\dB). Nothing in the model was set from them.
For the rigid hull, only the statement that the two full-wave solutions agree was known. The tile case (Section~\ref{sec:tile}) is not blind:
it was used to fix the conventions of the layer model, and the model was changed because of it.

\paragraph{Reference data.} The finite-element reflection coefficients and the lake measurement are published as figures only~\cite{yang2024}.
They were digitised from the figures, to about $\pm0.005$ in reflection coefficient and $\pm0.5$\dB{} in level.
The full-wave curves of the rigid hull were read from the vector paths of the figure in the arXiv version of~\cite{venas2020} and are exact to 0.02\dB.

\paragraph{Measure.} Levels are compared as energy means over aspect sectors, $10\lg\langle10^{\mathrm{TS}/10}\rangle$.
A sector mean is what a sonar sees of a target whose aspect it knows only roughly, and it is the quantity a model without phase between bodies can predict.
Aspect is measured from the bow: $0^\circ$ bow-on, $90^\circ$ on the beam, $180^\circ$ stern-on.

\section{Test 1: reflection of a rubber tile on plating}\label{sec:tile}

Yang et~al.~\cite{yang2024} give the plane-wave reflection coefficient of a homogeneous rubber tile at $0^\circ$, $45^\circ$, $60^\circ$ and $75^\circ$ incidence,
0.2--20\,kHz, computed with finite elements and periodic boundary conditions (their Fig.~7).
The tile is 50\,mm of rubber (1090\,kg/m$^3$, Young's modulus $30(1+0.249\,\mathrm i)$\,MPa, Poisson's ratio 0.49) on 8\,mm of steel with air behind.

\begin{figure}[!htb]
\centering
\includegraphics[width=\textwidth]{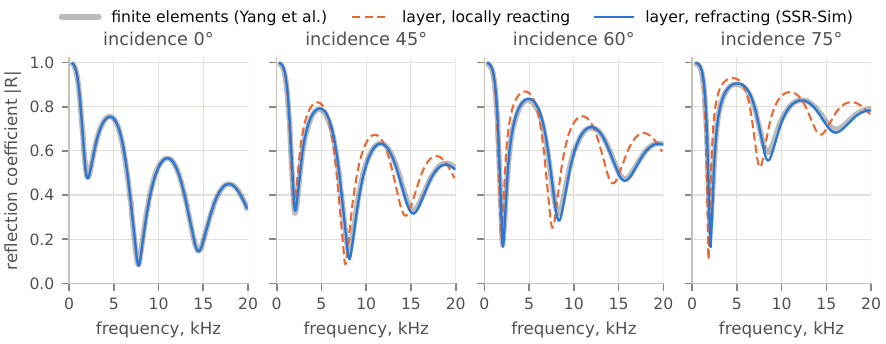}
\caption{Reflection coefficient of a 50\,mm rubber tile on 8\,mm steel with air behind. Grey: finite elements, digitised from~\cite{yang2024}.
Blue: the layer of Section~\ref{sec:beta}, refracting. Orange, dashed: the same layer taken as locally reacting.}\label{fig:tile}
\end{figure}

A one-dimensional layer responds through its plane-wave modulus,
\begin{equation}
  M=\frac{E(1-\nu)}{(1+\nu)(1-2\nu)} = 513\ \text{MPa},
\end{equation}
which gives a compressional speed of 686\,m/s in the rubber. Table~\ref{tab:tile} and Fig.~\ref{fig:tile} compare four versions of the layer model with the digitised curves.

\begin{table}[!htb]
\centering\small
\caption{Rubber tile on steel plating: difference between the layer model and the finite-element reflection coefficient of~\cite{yang2024},
0.3--20\,kHz, as rms / largest difference in $|R|$ (about 585 digitised points per angle).}\label{tab:tile}
\begin{tabular}{@{}lcccc@{}}
\toprule
Model & $0^\circ$ & $45^\circ$ & $60^\circ$ & $75^\circ$\\
\midrule
Refracting layer on the plate, air behind & 0.005 / 0.02 & 0.012 / 0.05 & 0.011 / 0.06 & 0.017 / 0.11 \\
The same, locally reacting & 0.005 / 0.02 & 0.078 / 0.17 & 0.102 / 0.21 & 0.091 / 0.34 \\
Locally reacting, rigid backing & 0.101 / 0.31 & 0.181 / 0.52 & 0.201 / 0.70 & 0.162 / 0.67 \\
Locally reacting, $E$ in place of $M$ & 0.373 / 0.85 & 0.352 / 0.70 & 0.311 / 0.67 & 0.207 / 0.65 \\
 
\bottomrule
\end{tabular}
\end{table}

Three things follow.
\begin{itemize}\setlength{\itemsep}{0pt}
\item At normal incidence the layer reproduces the finite-element curve to 0.005 rms, which is the accuracy of the digitisation.
This needs the plane-wave modulus, the loss factor as given, and the steel plate with air behind. With Young's modulus in place of $M$, or with a rigid backing, the curve is not reproduced.
\item At oblique incidence a locally reacting layer puts the resonances at the wrong frequencies and misses by 0.08--0.10 rms. With refraction in the layer the difference falls to 0.011--0.017 rms.
\item A free fit of the modulus and loss factor to all four panels returns $0.990\,M$ and $\eta=0.255$. The published curves are therefore the one-dimensional layer solution, and the periodic finite-element model adds nothing to it for a homogeneous tile.
\end{itemize}
SSR-Sim's layer was locally reacting before this test. The refracting form was added because of it and is used in the engineering profile from here on.

\section{Test 2: the rigid BeTSSi submarine at 1\,kHz (blind)}\label{sec:rigid}

\paragraph{Reference.} Ven{\aa}s and Kvamsdal~\cite{venas2020} give the monostatic TS of the rigid BeTSSi submarine at 1\,kHz in the horizontal plane (their Fig.~36),
computed with an isogeometric boundary-element method on two meshes and compared with a boundary-element result from WTD~71.
Their two meshes agree to 0.03\dB{} rms. The two codes agree to 1.6\dB{} rms point by point, to 0.4\dB{} rms once smoothed over $5^\circ$, and to 0.2\dB{} in every sector mean (Table~\ref{tab:rigid}).
The reference is therefore known to a few tenths of a decibel at the level of this comparison.

\paragraph{Case.} The geometry is the improved BeTSSi hull of~\cite{venas2020}: an ellipsoidal bow with semi-axes 7 and 3.5\,m, a 42\,m parallel body of radius 3.5\,m
with a flat deck above, a tail of 13\,m, a sail and rudders with NACA sections; 62\,m in all.
At 1\,kHz the wavelength is 1.5\,m and $kb=14.7$ on the hull radius, which is near the lower limit of a high-frequency model.
The hull was cut as in Section~\ref{sec:cut}, and the sail and the vertical rudders were modelled as in Section~\ref{sec:sail}.
Table~\ref{tab:bodies} and Fig.~\ref{fig:bodies} give the nine bodies.

The lower hull is a body of revolution up to 1.75\,m above the axis, so in the horizontal plane the hull is a spheroid, a cylinder, a curved transition and a frustum.
The deck and the panels that join it to the hull have no horizontal normal and are left out.
The upper and lower vertical rudders are at the same range and are taken as one body of their combined exposed height.
Their curved fronts fail the rule $k\rho\ge1$ at aspects below $59.5^\circ$ and are not counted there.
The horizontal rudders and the depth rudders have no specular area in the horizontal plane and are left out, as is every edge and junction.
The surface is rigid, the range is the far field, and the computation takes 0.2\,s for 1801 aspects.

\begin{table}[!htb]
\centering\small
\caption{Bodies of the rigid BeTSSi case (full scale), with the forms of Table~\ref{tab:forms}. $\rho$ is the radius of curvature in plan at the specular point.}\label{tab:bodies}
\begin{tabular}{@{}rlp{6.9cm}l@{}}
\toprule
 & Body & Form and dimensions & Aspects\\
\midrule
1 & Bow & half prolate spheroid, semi-axes 7.0 and 3.5\,m & $0$--$90^\circ$\\
2 & Parallel body & cylinder, length 42\,m, radius 3.5\,m & flash at $90^\circ$\\
3 & Tail transition & doubly curved surface: meridian radius 21.0\,m, transverse radius 3.5 to 2.6\,m & $90$--$108^\circ$\\
4 & Tail cone & frustum, half-angle $18^\circ$, radius 2.47 to 0.36\,m, length 6.5\,m & flash at $108^\circ$\\
5 & Tail end & disc, radius 0.36\,m & normal at $180^\circ$\\
6 & Sail, front & vertical cylinder, height 3.5\,m, $\rho$ of the NACA section (chord 12.65\,m, width 2.2\,m): 0.42\,m at the leading edge, 19.3\,m abeam & $0$--$90^\circ$\\
7 & Sail, sides & plates $8.92\times3.5$\,m & flash at $97.1^\circ$\\
8 & Vertical rudders, front & as 6: chord 2.44\,m, width 0.34\,m, combined height 5.11\,m & $59.5$--$90^\circ$\\
9 & Vertical rudders, sides & plates $1.72\times5.11$\,m & flash at $95.6^\circ$\\
\bottomrule
\end{tabular}
\end{table}

\begin{figure}[!htb]
\centering
\includegraphics[width=\textwidth]{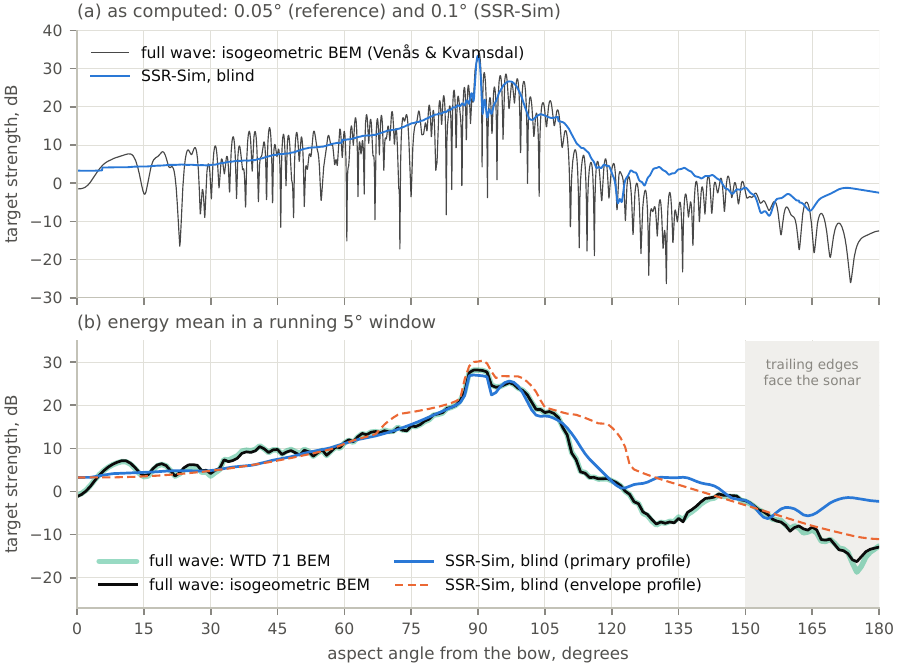}
\caption{Rigid BeTSSi submarine at 1\,kHz, monostatic, horizontal plane. (a) The full-wave reference and the blind prediction as computed.
(b) The same after an energy mean in a running $5^\circ$ window, with the second full-wave solution and the envelope profile.
Reference curves are from Fig.~36 of~\cite{venas2020}.}\label{fig:rigid}
\end{figure}

\paragraph{Result.} Table~\ref{tab:rigid} gives the sector levels and Fig.~\ref{fig:rigid} the curves.

\begin{table}[!htb]
\centering\small
\caption{Rigid BeTSSi submarine, 1\,kHz: sector levels of the full-wave reference~\cite{venas2020} and differences from it, dB. The prediction is blind.}\label{tab:rigid}
\begin{tabular}{@{}lrrrrrrr@{}}
\toprule
Aspect sector, degrees & 0--30 & 30--75 & 75--105 & 105--150 & 150--180 & 0--150 & 0--180\\
\midrule
Isogeometric BEM (reference), dB & 5.0 & 11.1 & 23.6 & 8.6 & $-$8.2 & 17.2 & 16.4 \\
WTD 71 BEM $-$ reference & $-$0.2 & $+$0.1 & 0.0 & 0.0 & $-$0.1 & 0.0 & 0.0 \\
\addlinespace
SSR-Sim, primary profile $-$ reference & $-$0.7 & $-$0.5 & $-$0.6 & $+$0.5 & $+$4.9 & $-$0.6 & $-$0.6 \\
SSR-Sim, envelope profile $-$ reference & $-$1.3 & $+$0.8 & $+$2.2 & $+$4.7 & $+$1.2 & $+$2.2 & $+$2.2 \\
 
\bottomrule
\end{tabular}
\end{table}

From the bow to $150^\circ$ every sector mean of the prediction is within 0.7\dB{} of the full-wave result, and the mean over $0$--$150^\circ$ is 0.6\dB{} low.
The beam flash peaks at 33.4\dB{} against 34.1\dB. The model also places a second, broader flash 7\dB{} lower at $97^\circ$, where the flat sides of the sail face the sonar.
The reference has it: smoothed over $5^\circ$, the two are 25.6 and 25.3\dB{} there.

The smoothed curves show where the agreement is local and where it holds only on average. Over $0$--$150^\circ$ the prediction is within 3\dB{} of the reference at 79\% of the angles, with an rms difference of 3.2\dB.
It is 3\dB{} low near $10^\circ$ and $40^\circ$, where the reference has broad lobes from the interference of the bow and the sail that an energy sum cannot hold.
It is up to 10.7\dB{} high in a notch of the reference near $130^\circ$.

\begin{figure}[!htb]
\centering
\includegraphics[width=\textwidth]{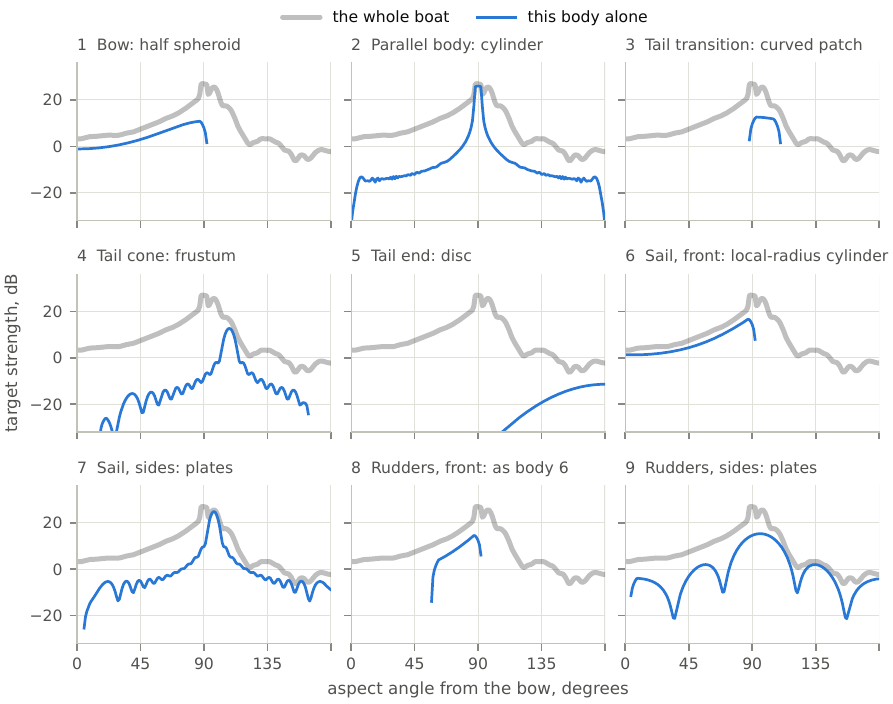}
\caption{Rigid BeTSSi submarine at 1\,kHz: the return of each body of Table~\ref{tab:bodies} in the blind prediction (blue) against that of the whole boat (grey),
as target strength after an energy mean in a running $5^\circ$ window. A curved body returns only within its range of normals; a ruled body has a flash and side lobes.}\label{fig:parts}
\end{figure}

\paragraph{What each body returns.} Table~\ref{tab:shares} and Fig.~\ref{fig:parts} split the prediction by body.
Forward of the beam the echo is the curved front of the sail and the bow, in that order: the front of the sail carries about half the energy and the bow a third.
In the beam sector the parallel body carries 36\% and the flat sides of the sail 32\%, in a flash of their own at $97^\circ$.
Aft of the beam ($105$--$150^\circ$) the tail cone, with its flash at $108^\circ$, carries a third, the curved run ahead of it 12\%, and the side lobes of the sail and rudder plates the remaining half.
Astern, those side lobes are four fifths of the predicted energy.

\begin{table}[!htb]
\centering\small
\caption{Rigid BeTSSi submarine, 1\,kHz: share of each body in the echo energy of each aspect sector in the blind prediction (primary profile), per cent. A dash is less than 0.5\%.}\label{tab:shares}
\begin{tabular}{@{}rlrrrrr@{}}
\toprule
 & Aspect sector, degrees & 0--30 & 30--75 & 75--105 & 105--150 & 150--180\\
\midrule
1 & Bow & 33 & 31 & 3 & -- & -- \\
2 & Parallel body & 1 & 1 & 36 & 2 & 7 \\
3 & Tail transition & -- & -- & 4 & 12 & -- \\
4 & Tail cone & -- & -- & -- & 33 & 2 \\
5 & Tail end & -- & -- & -- & -- & 12 \\
\addlinespace
6 & Sail, front & 53 & 46 & 8 & -- & -- \\
7 & Sail, sides & 4 & 3 & 32 & 14 & 42 \\
8 & Vertical rudders, front & -- & 12 & 5 & -- & -- \\
9 & Vertical rudders, sides & 9 & 6 & 11 & 39 & 37 \\
 
\bottomrule
\end{tabular}
\end{table}

In the stern sector the prediction is 5\dB{} high. Nothing there faces the sonar except the small end disc. The model's level comes from the far side lobes of the sail and rudder plates,
which stand for the returns of their ends. Seen from astern those ends are the trailing edges, and a lobe function is a poor model of a sharp edge at grazing incidence.
The envelope profile happens to be closer in this sector ($+1.2$\dB) and is worse elsewhere: 2.2\dB{} high over $0$--$150^\circ$ and 4.7\dB{} high aft of the beam.

\section{Test 3: a tile-coated 1:15 scale model in a lake (blind)}\label{sec:scale}

\paragraph{Reference.} Yang et~al.~\cite{yang2024} measured the monostatic TS of a 1:15 BeTSSi model in a lake at 5, 10, 15 and 20\,kHz over all aspects (their Fig.~12),
and computed it with a Kirchhoff facet model using finite-element reflection coefficients.
The model is 4.13\,m long with a 3\,mm shell and a midship radius of 0.25\,m. It is covered with homogeneous tiles 10\,mm thick (1100\,kg/m$^3$, Young's modulus 4.41\,MPa, Poisson's ratio 0.48).
Source, hydrophone and model were in line, the hydrophone 3.65\,m from the model and the source 7.15\,m beyond it.
The authors attribute the weak agreement of their own computation over the stern aspects to the processing and suspension of the model.

\paragraph{Case.} The bodies are those of the source paper's drawing, which gives overall dimensions only (Table~\ref{tab:bodies3}).
The drawing has no section for the sail, so the simpler variant of Section~\ref{sec:sail} was used: a half-ellipse in front, cut into four sectors, and flat tapered sides.
The vertical fins are one plate, and the curved run at the stern is not separated from the cone.
The shell is 3\,mm steel (7800\,kg/m$^3$) with the tile on every surface and water inside (1000\,kg/m$^3$, 1500\,m/s).
The tile's plane-wave modulus is 38.7\,MPa, following Test~1. Its loss factor is not given; 0.249, the value of the same paper's other rubber, was assumed, with 0.10--0.40 as a band.

With source and receiver in line at 10.8 and 3.65\,m the Fresnel test of Section~\ref{sec:range} takes the range $2r_{\mathrm s}r_{\mathrm r}/(r_{\mathrm s}+r_{\mathrm r})=5.46$\,m,
at which $2\rhoF$ is 1.8\,m at 5\,kHz and 0.9\,m at 20\,kHz.
The parallel body, 2.97\,m long at model scale, is in its near field at every frequency and takes the long form; every other ruled body is short.

\begin{table}[!htb]
\centering\small
\caption{Bodies of the scale-model case. Dimensions are at full scale; the model is 1:15.}\label{tab:bodies3}
\begin{tabular}{@{}lp{9.3cm}l@{}}
\toprule
Body & Form and dimensions & Aspects\\
\midrule
Bow & half prolate spheroid, semi-axes 7.0 and 3.75\,m & $0$--$90^\circ$\\
Parallel body & cylinder, length 44.6\,m, radius 3.75\,m & flash at $90^\circ$\\
Stern cone & frustum, half-angle $18^\circ$, radius 3.75 to 0.53\,m, length 9.9\,m & flash at $108^\circ$\\
Sail, front & half-ellipse $3.85\times1.1$\,m as four vertical cylinders of height 3.5\,m: radius 0.35, 0.79, 3.82 and 11.9\,m for aspects $0$--$30^\circ$, $30$--$60^\circ$, $60$--$80^\circ$ and $80$--$90^\circ$ & $0$--$90^\circ$\\
Sail, sides & plates $8.92\times3.5$\,m & flash at $97.1^\circ$\\
Vertical fins & one plate $2.5\times5.12$\,m & flash at $90^\circ$\\
\bottomrule
\end{tabular}
\end{table}

Two things were decided before the comparison. First, aspects $120$--$180^\circ$ are reported but not counted, because of the suspension.
Second, 5, 15 and 20\,kHz are the test and 10\,kHz is an observation. At 10\,kHz the tile is half a wavelength thick and its reflection depends strongly on the assumed loss:
the band of loss factors moves the prediction by 4\dB{} there and by about 1\dB{} at the other three frequencies.

\begin{table}[!htb]
\centering\small
\caption{Tile-coated 1:15 scale model: measured sector levels (digitised from~\cite{yang2024}) and differences from them, dB.
SSR-Sim is the blind prediction; Kirchhoff--FE is the computation of~\cite{yang2024}. The last column is affected by the model's suspension and is not part of the test.}\label{tab:scale}
\begin{tabular}{@{}rlrrrrrr@{}}
\toprule
kHz & Aspect sector, degrees & 0--30 & 30--75 & 75--105 & 105--120 & 0--120 & 120--180\\
\midrule
5 & Measured, dB & $-$14.0 & $-$9.4 & $-$3.8 & $-$9.9 & $-$8.1 & $-$14.2 \\
 & SSR-Sim $-$ measured & $-$7.4 & $-$5.1 & $-$3.1 & $-$0.7 & $-$3.1 & $-$9.2 \\
 & Kirchhoff--FE $-$ measured & $-$4.8 & $-$4.3 & $-$3.2 & $-$6.0 & $-$3.0 & $-$8.3 \\
\addlinespace
10 & Measured, dB & $-$11.6 & $-$7.1 & $-$3.7 & $-$7.6 & $-$7.1 & $-$13.0 \\
 & SSR-Sim $-$ measured & $-$12.8 & $-$11.3 & $-$5.6 & $-$7.7 & $-$6.9 & $-$14.0 \\
 & Kirchhoff--FE $-$ measured & $-$7.0 & $-$5.0 & $-$1.3 & $-$10.9 & $-$2.4 & $-$7.0 \\
\addlinespace
15 & Measured, dB & $-$12.2 & $-$11.4 & $-$7.5 & $-$10.7 & $-$10.1 & $-$14.7 \\
 & SSR-Sim $-$ measured & $-$7.8 & $-$2.3 & $+$3.8 & $-$0.3 & $+$1.4 & $-$9.9 \\
 & Kirchhoff--FE $-$ measured & $-$4.2 & $+$0.5 & $+$5.3 & $+$2.1 & $+$3.1 & $-$9.9 \\
\addlinespace
20 & Measured, dB & $-$12.5 & $-$11.1 & $-$5.5 & $-$8.6 & $-$9.0 & $-$12.2 \\
 & SSR-Sim $-$ measured & $-$8.4 & $-$3.8 & $+$1.1 & $-$4.4 & $-$0.5 & $-$13.1 \\
 & Kirchhoff--FE $-$ measured & $-$3.9 & $+$1.7 & $+$4.6 & $-$0.1 & $+$3.5 & $-$10.7 \\
 
\bottomrule
\end{tabular}
\end{table}

\begin{figure}[!htb]
\centering
\includegraphics[width=\textwidth]{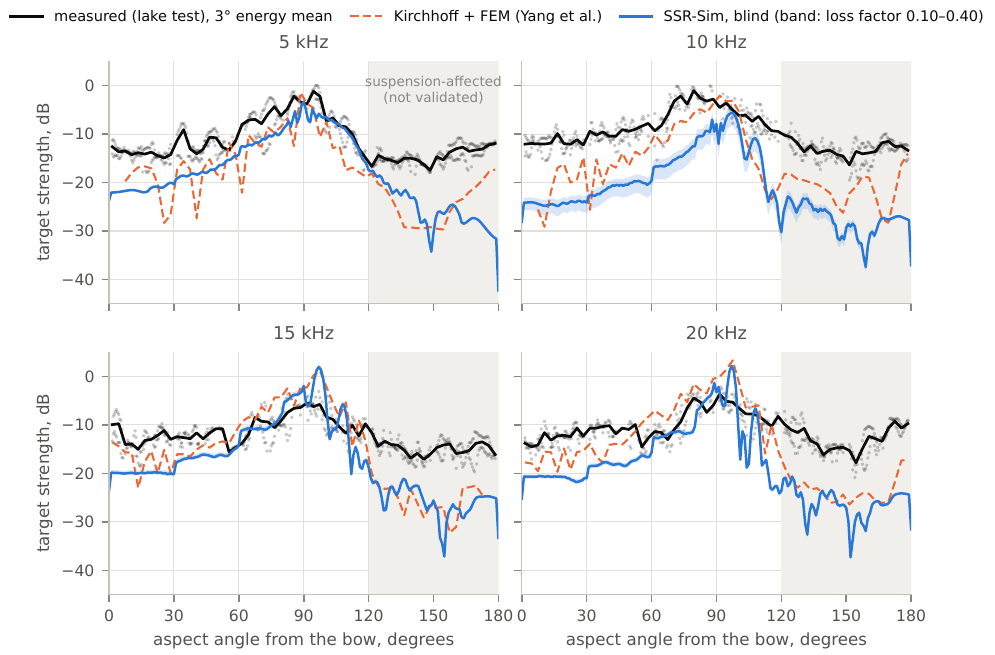}
\caption{Tile-coated 1:15 BeTSSi model: measured TS (points: digitised values from both sides of the model; black line: their energy mean in $3^\circ$ bins),
the Kirchhoff--finite-element computation of~\cite{yang2024}, and the blind SSR-Sim prediction with its band for tile loss factors 0.10--0.40.}\label{fig:scale}
\end{figure}

\paragraph{Result.} Table~\ref{tab:scale} and Fig.~\ref{fig:scale} give the comparison.
\begin{itemize}\setlength{\itemsep}{0pt}
\item Over $0$--$120^\circ$ the prediction is within $-3.1$, $+1.4$ and $-0.5$\dB{} of the measurement at 5, 15 and 20\,kHz.
The Kirchhoff--finite-element computation is within $-3.1$, $+3.1$ and $+3.5$\dB.
\item Near the beam ($75$--$120^\circ$) the prediction is within $-4.4$ to $+3.8$\dB.
\item In the bow sector the prediction is 7.4--8.4\dB{} low at the three test frequencies, and 2.3--5.1\dB{} low at $30$--$75^\circ$.
The measurement has a floor of about $-12$ to $-15$\dB{} at every aspect away from the beam and at every frequency. Neither model has it; the Kirchhoff computation is 3.9--4.8\dB{} low at the bow.
\item At 10\,kHz the prediction is 5.6\dB{} low at the beam and 6.9\dB{} low over $0$--$120^\circ$.
The measured beam level is the same at 10\,kHz as at 5\,kHz ($-3.7$ and $-3.8$\dB), so the measurement shows no resonance of the tile there.
Across the band of loss factors the difference over $0$--$120^\circ$ runs from $-9.6$ to $-5.5$\dB: the data favour more loss than was assumed, or a resonance elsewhere.
\item Over the stern aspects the prediction is 9--14\dB{} low and the Kirchhoff computation 7--11\dB{} low.
\end{itemize}

\paragraph{Sensitivity.} With air inside the model the result changes by 0.1\dB{} at 5, 15 and 20\,kHz and by 0.4--0.8\dB{} at 10\,kHz.
Taking the far field in place of the test geometry raises the beam sector by 2.0--3.4\dB{} and the level over $0$--$120^\circ$ by 1.7--2.5\dB,
which improves 5\,kHz and worsens 15 and 20\,kHz.
The envelope profile gives $-0.9$, $+3.9$ and $+2.0$\dB{} over $0$--$120^\circ$ at 5, 15 and 20\,kHz.

\section{Discussion}

\paragraph{What the tests establish.} The layer model reproduces an independent finite-element solution of a tile on plating.
For a rigid hull near the lower limit of its frequency range, the component model gives sector levels within 1\dB{} of full-wave results from the bow to $150^\circ$,
from nine bodies and 0.1\,ms per aspect.
Against a real measurement of a coated model its level over aspects $0$--$120^\circ$ is within about 3\dB{} at three frequencies out of four.
That is no worse than a facet Kirchhoff computation with finite-element coefficients, and at 15 and 20\,kHz it is closer.

The rigid case shows why so small a model can work. Away from the stern, the echo of this hull at this frequency is a handful of specular returns,
and the mean level of each is set by the local curvature and extent of one surface. Phase between the returns moves the lobes and leaves the sector means alone.
The sail matters as much as the hull: its curved front carries half the echo energy forward of the beam, and its flat sides carry nearly as much energy in the beam sector as the hull itself.

\paragraph{Where it fails.} The failures are of three kinds.

\emph{Edges.} The model has no diffraction term. In the rigid case the stern sector is 5\dB{} high because the side lobes of plates stand in for trailing-edge returns.
An edge term with the correct behaviour at grazing incidence is the first extension the comparison calls for~\cite{fawcett2001}.

\emph{The off-beam floor of the measured model.} The lake measurement does not fall below about $-15$\dB{} at any aspect, and the forward-sector deficit of 7--8\dB{} follows from that.
The rigid full-wave comparison has no such deficit: its bow sector is predicted to 0.7\dB. The missing energy is therefore not in the specular physics of the hull form.
Candidates are the elastic response of a 3\,mm shell, the joints between tiles, internal structure, the suspension gear, whose echoes the authors believe they see in their time records, and the noise floor of the measurement.
The Kirchhoff model of the source paper shares the deficit, which points the same way.

\emph{A resonant coating.} At 10\,kHz the prediction rests on a tile resonance fixed by a static modulus and an assumed loss, and it is 7\dB{} low.
A coating model is only as good as the dynamic properties it is given.

\paragraph{Limits of the tests.} The tests cover the specular component and the reflection coefficient of a coated plate.
The model's terms for the structure between two hulls and for stiffened plating are untested here; no open benchmark exists for them.
The rigid case is one frequency in one plane. The measured case is one model in one trial, read from figures, with an unknown medium inside and an assumed loss factor.
At model scale the sail's leading edge is below $k\rho=1$ at 5\,kHz, and the rule applied in the rigid case was not applied there.
The model does not hide one body behind another: seen from ahead, the vertical rudders of the rigid case lie behind the hull, and the side lobes of their plates still supply 9\% of the energy of the bow sector.
The change from the short to the long form at $L=2\rhoF$ is a step, where the true transition is smooth.
The cutting of the hull into bodies was done by the author of the model. Another user would cut it differently, and the spread between users is not measured here.

\section{Conclusions}

\begin{enumerate}\setlength{\itemsep}{0pt}
\item A homogeneous tile on plating is described at all incidence angles by a refracting fluid layer with the plane-wave modulus. A locally reacting layer fails at oblique incidence.
\item A blind prediction of the rigid BeTSSi submarine at 1\,kHz from nine simple bodies agrees with full-wave results to 0.7\dB{} in every sector from the bow to $150^\circ$, and is 5\dB{} high astern.
\item A sail needs two bodies: a vertical cylinder with the local radius of curvature of its section for the front, and a flat plate for each side.
On the rigid benchmark the front carries half the echo energy forward of the beam, and the sector levels there are predicted to 0.7\dB.
The side lobes of the plates are a poor stand-in for the trailing edge and account for the error astern.
\item A blind prediction of a tile-coated scale model agrees with a lake measurement to about 3\dB{} over aspects $0$--$120^\circ$ at 5, 15 and 20\,kHz. It is 7\dB{} low on a tile resonance and 7--8\dB{} low in the bow sector, where the measurement has a floor that specular models do not reproduce.
\item The next steps are an edge-diffraction term, a test of the rigid hull at more frequencies and elevations, and open data for double hulls.
\end{enumerate}

\section*{Acknowledgements and declarations}

The author is the developer of SSR-Sim. The work received no external funding.
The benchmark scripts and the text were prepared with the assistance of an AI model (Claude, Anthropic).
The author thanks the authors of~\cite{yang2024} and~\cite{venas2020} for publishing a measurement and a benchmark that others can use.

\section*{Data}

The program is not distributed. Section~\ref{sec:model} with Tables~\ref{tab:bodies} and~\ref{tab:bodies3} is meant to be sufficient to recompute the predictions.
As a check, two short scripts that use only the formulas of Section~\ref{sec:model} recompute the two hull cases and reproduce the recorded predictions to the last digit written.
The prediction files named in Appendix~\ref{app:hash} and the digitised reference data of Tests~1 and~3 are published under CC BY 4.0~\cite{tyagi2026data} at
\href{https://doi.org/10.57967/hf/11704}{doi:10.57967/hf/11704}, with a script that checks the two files against the hashes of Appendix~\ref{app:hash}.
The recomputation scripts and the comparison scripts are available from the author.

\appendix
\section{Recorded predictions}\label{app:hash}

\begin{table}[!htb]
\centering\small
\begin{tabular}{@{}p{4.3cm}p{2.9cm}p{7.6cm}@{}}
\toprule
File & Recorded (UTC) & SHA-256\\
\midrule
Scale model, prediction & 18 Nov 2026, 17:05 & \texttt{a7788ce8c98434e764f802a6c6aa8976 ef6188653da2f64ece4771a3e9584a06}\\
Rigid hull, prediction & 18 Nov 2026, 18:20 & \texttt{8f41ce1c80b02037ad1b12a3a9f98611 fde07f9bb6f1e23c4474ad9dcc963787}\\
SSR-Sim r10, source & 18 Nov 2026 & \texttt{f45b0945b1d58add951dfc1a377d3e31 bbdd01aa32a8e2e39ebeffeb49284841}\\
\bottomrule
\end{tabular}
\end{table}

\noindent The recorded levels are $20\lg\Req-6.0$, which is 0.02\dB{} above eq.~(1). Each prediction file holds both profiles. The scale-model file also holds the band of loss factors, water and air inside, and the test range and the far field.

\noindent The two prediction files are published under these names~\cite{tyagi2026data}. The data set, \href{https://doi.org/10.57967/hf/11704}{doi:10.57967/hf/11704},
includes the script \texttt{verify\_hashes\_r1.py}, which checks them against the first two hashes above.

\end{document}